\documentclass[prb,twocolumn,showpacs,preprintnumbers,amsmath,amssymb]{revtex4}
\usepackage{graphicx}% Include figure files

\usepackage{dcolumn}% Align table columns on decimal point
\usepackage{bm}% bold math

\begin{document}

\newcommand*{\cm}{cm$^{-1}$\,}

%\preprint{APS/123-QED}

\title{Optical spectroscopy study on CeTe$_3$: evidence for multiple charge-density-wave orders}
\author{B. F. Hu}
\author{P. Zheng}
\author{R. H. Yuan}
\author{T. Dong}
\author{B. Cheng}
\author{Z. G. Chen}
\author{N. L. Wang}

\affiliation{Beijing National Laboratory for Condensed Matter
Physics, Institute of Physics, Chinese Academy of Sciences,
Beijing 100080, People's Republic of China}
%


\begin{abstract}
We performed optical spectroscopy measurement on single crystal of
CeTe$_3$, a rare-earth element tri-telluride charge density wave
(CDW) compound. The optical spectra are found to display very
strong temperature dependence. Besides a large and pronounced CDW
energy gap being present already at room temperature as observed
in earlier studies, the present measurement revealed the formation
of another energy gap at smaller energy scale at low temperature.
The second CDW gap removes the electrons near E$_F$ which undergo
stronger scattering. The study yields evidence for the presence of
multiple CDW orders or strong fluctuations in the light rare-earth
element tri-telluride.
\end{abstract}

\pacs{78.30.Er, 78.40.Kc, 75.50.Cc}

\maketitle

\section{Introduction}
Collective quantum phenomena, such as charge density wave (CDW) and
spin density wave (SDW) , are among the most fascinating phenomena
in solids and have been a subject of considerable interest in modern
condensed matter physics. Most of CDW or SDW states are originated
from the nesting topology of Fermi surfaces. This results in a
divergence or strong anomaly in susceptibility at the nesting wave
vector, leading to an instability of the electronic structure. The
CDW or SDW states are stabilized via electron-phonon or
electron-electron interactions by opening up energy gaps in the
nested regions of the Fermi surfaces, which thus leads to a lowering
of the electronic energies of the occupied states. Formation of an
energy gap at the transition has been generally taken as a
characteristic feature of a CDW or SDW order.

Among various CDW materials, the rare-earth tri-telluride RTe$_3$
(R=Y, La, and rare earth elements) has attracted much
attention\cite{TEMCDW} due to their widely tunable properties by
either chemical
substitutions\cite{2CDW,RTe3che.pres,ARPESRTe3,quenchCDW} or
application of
pressure\cite{CeTe3pres,RTe3pres,CeTe3mulphase,quenchCDW}. RTe$_3$
has a layered structure\cite{structure,structure2} (inset of Fig.
1), consisting of the alternate stacking of insulating corrugated
RTe slabs and two square Te-layers along the b-axis. The crystal
lattice is weakly orthorhombic, which belongs to the space group
\emph{Cmcm}(No. 63)\cite{structure}.Under much high pressure, the
difference between the two short axes gradually disappears and the
structure seems to undergo a transition from orthorhombic to
tetragonal\cite{quenchCDW}. R in the compound is trivalent, donating
three electrons to the system. They completely fill the Te p
orbitals in the RTe slabs, but partially those Te p orbitals in the
square Te-layers\cite{structure2,ARPESCeTe3400meV}. Metallic
conduction occurs in the Te layers, leading to highly anisotropic
transport properties\cite{structure2,material}.

Band structure calculations indicate two very simple Fermi
surfaces\cite{theory,ARPESRTe3} which exhibit little dispersion
along b axis. It is revealed by
ARPES\cite{ARPESCeTe3400meV,ARPESCeTe3} that the nesting of the
Fermi surfaces drives the CDW instability. In the family the
incommensurate nesting wave vector is about much the same which is
along the \emph{c}$^{\ast}$ in the base plane of the reciprocal
space\cite{TEMCDW,SmTe3}. Angle resolved photon emission
spectroscopy measurement demonstrated the gap scale distribution in
the first brillouin zone, which reaches the maximum in the optimal
nested region at \emph{k}$_x$=0 and decreases to zero far from
\emph{a$^\ast$} axis\cite{ARPESRTe3}. The maximum gap value
decreases from light to heavy rare-earth element. For CeTe$_3$, a
very large energy gap ($\approx$ 400 meV)\cite{ARPESCeTe3400meV} has
been revealed by ARPES. It is expected by the mean field theory that
the CDW transition should appear even above the melting temperature
of the compound\cite{SmTe3}. However, the CDW ordering temperatures
are found to be substantially reduced from the light to heavier rare
earth elements\cite{2CDW}.

Optical spectroscopy is a powerful technique to probe the energy
gap in the ordered state. It also yields fruitful information
about conducting carrier response. Optical measurements on
CeTe$_3$ at room temperature\cite{RTe3che.pres} and under
pressure\cite{RTe3pres,CeTe3pres} have been reported, which
provided clear evidence for the formation of the partial energy
gap\cite{ARPESCeTe3400meV} in CDW state and its evolution with
pressure. Since the CDW transition temperature of this compound is
extremely high\cite{2CDW}, the room temperature is believed to be
already deeply into the CDW state, and it is claimed that no
temperature dependence exists in the optical spectra below room
temperature\cite{RTe3che.pres}. Here we present a temperature
dependent optical measurement on CeTe$_3$. In contrast to the
early assertion, we observed a prominent temperature dependence of
the optical spectra. Much more surprisingly, our measurements
revealed a new gap structure developing below 300 K at lower
energy scale, evidencing the formation of another CDW order at low
temperature.

\section{\label{sec:level2}EXPERIMENT AND RESULTS}
The single crystals of CeTe$_3$ were grown by a self-flux
method\cite{material}. Ce and Te elements with the molar ratio of
1:40 were mixed, then put into a alumina crucible and sealed in a
quartz tube under vacuum. The mixture was heated to 860 $^0$C and
stayed for 10 hours, then slowly cooled to 560 $^0$C at a rate of 3
$^0$C/h. At the end temperature the rest flux Te is still in liquid
and separated from the crystals in a centrifuge. Plate-like single
crystals with the dimension of 3$\times$3 mm$^2$ were obtained after
breaking the crucible. The samples are stored in a glove-box under
argon atmosphere since the compound is somewhat air and moisture
sensitive. The crystals were characterized by the x-ray diffraction
(XRD) and scanning electron microscopy (SEM) measurements. Figure 1
shows the \textit{(0k0)} XRD patterns for the single crystal samples
of CeTe$_3$ with Cu K$\alpha$ radiation. The XRD patterns indicate
that the samples are of the characteristic of good crystallization
along the b axis. The obtained b-axis lattice parameter is b=26.0348
$\AA$, which is consistent with reported lattice parameter in
literature\cite{quenchCDW}. The energy dispersive spectroscopy (EDS)
analysis equipped with the SEM indicates the correct Ce:Te=1:3
ratio.

\begin{figure}[t]
\scalebox{0.40} {\includegraphics [bb=370 95 8cm 18cm]{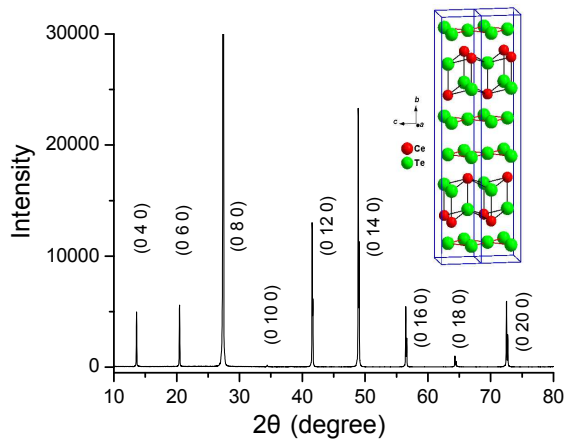}}
\caption{(Color online) The \textit{(0k0)} x-ray diffraction pattern
of single crystal CeTe$_3$. The strongest peak(intensity up to near
10$^5$) is only partially displayed in order to show other peaks
clearly. Inset shows the crystal structure.}
\end{figure}

The temperature dependent in-plane (ac-plane) resistivity was
obtained by the four contact technique in a Quantum Design physical
properties measurement system (PPMS) and plotted in figure 2. The
resistivity shows good metallic behavior below room temperature. In
agreement with previous studies\cite{material}, a sharp drop near 3
K is observed, which could be attributed to the antiferromagnetic
ordering of the spins from the localized 4f electrons of
Ce\cite{material,CeTe3mulphase,magRTe3}. The above characterizations
indicate a good quality of the single crystal.

\begin{figure}[b]
\includegraphics[width=6.5cm,clip]{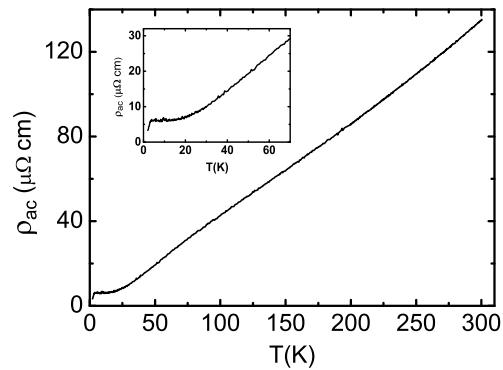}

%\scalebox{0.29} {\includegraphics [bb=520 20 8cm
%18cm]{resis1.eps}}
\caption{(Color online) The temperature
dependent in-plane (ac-plane) resistivity of single crystal
CeTe$_3$. Inset: the expanded plot of $\rho$(T) in the low
temperature range.}
\end{figure}

The near-normal incident reflectance spectra were measured by
Bruker IFS 113v and 66v/s spectrometers in the frequency range
from 40 to 25000 cm$^{-1}$. An \textit{in situ} gold and aluminium
overcoating technique was used to get the reflectivity
R($\omega$). The real part of conductivity $\sigma_1(\omega)$ is
obtained by the Kramers-Kronig transformation of R($\omega$).  A
Hagen-Rubens relation was used for low frequency extrapolation. A
constant value was used for high frequency extrapolation up to
100000 cm$^{-1}$, above which a $\omega^{-4}$ dependence is
employed.

Figure 3 shows the R($\omega$) spectra up to 6500 \cm with its
inset revealing the expanded range up to 25000 \cm. Meanwhile, the
real part of conductivity $\sigma_1(\omega)$ is displayed in
figure 4. At room temperature, the reflectance R($\omega$) shows
rather high values at low frequency, but decreases rapidly with
increasing frequency. A pronounced dip is seen near 3500 \cm. This
leads to a strong peak in the $\sigma_1(\omega)$, roughly near
4600 \cm. The spectra provides optical evidence for the presence
of an energy gap at room temperature\cite{2CDW}. The room
temperature data are consistent with those reported in earlier
studies\cite{RTe3che.pres}. Besides this peak in the mid-infrared
region, there exists a very sharp Drude component centered at zero
frequency. The spectra demonstrate that the compound remains
metallic even it is in the CDW state. This is also clearly
illustrated in the above resistivity measurement\cite{material}.
Obviously, the Fermi surfaces are only partially gapped in the CDW
state\cite{ARPESCeTe3400meV,ARPESCeTe3,ARPESRTe3}.

\begin{figure}[b]
\scalebox{0.3} {\includegraphics [bb=520 20 8cm 18cm]{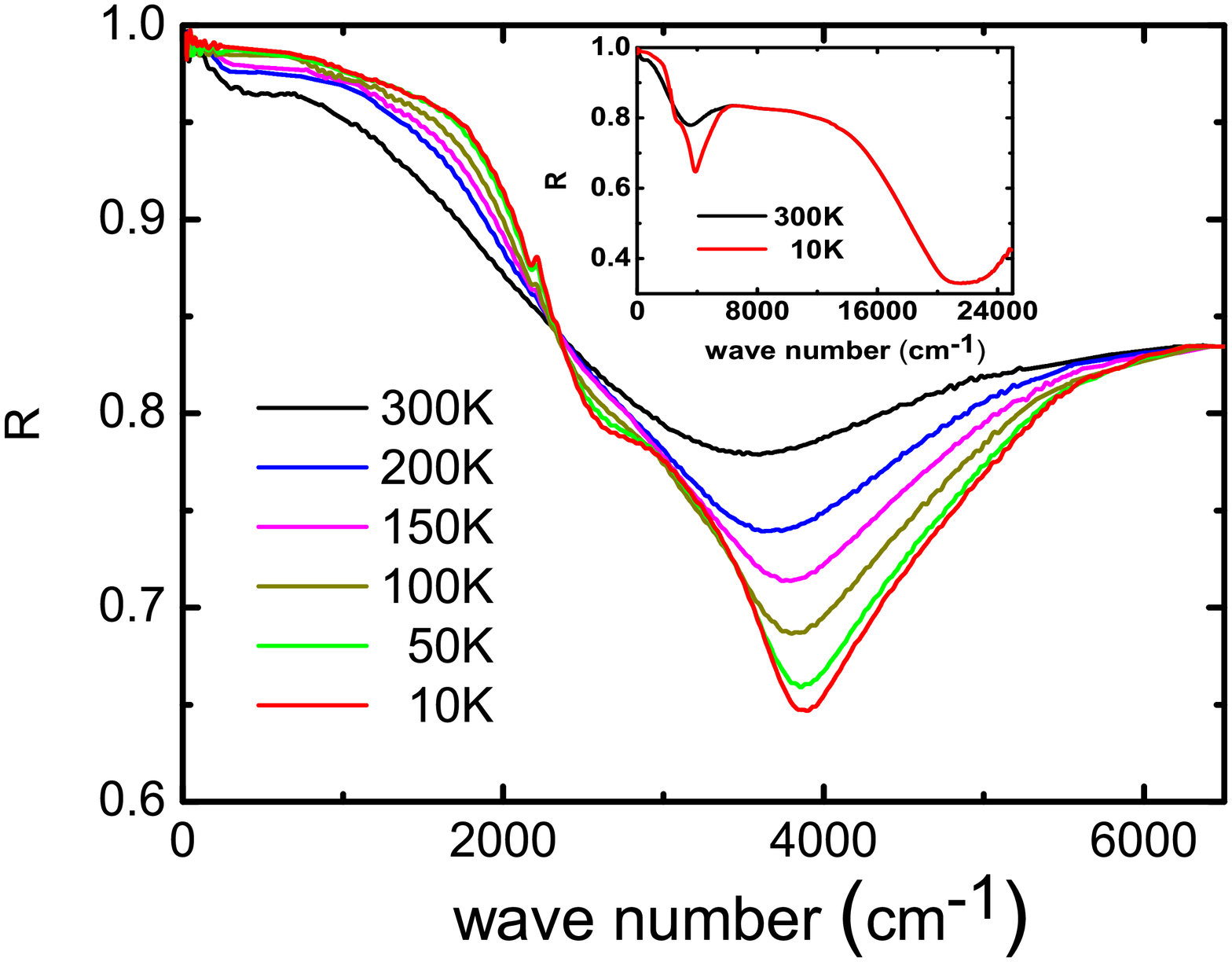}}
\caption{(Color online) The temperature dependent reflectivity of
CeTe$_3$ in the range from 40 to 6500 \cm. Inset shows the
R($\omega$) at two representative temperatures over a broad
frequency range from 40 to 25000 \cm.}
\end{figure}

\begin{figure}[t]
\scalebox{0.3} {\includegraphics [bb=520 20 8cm 19cm]{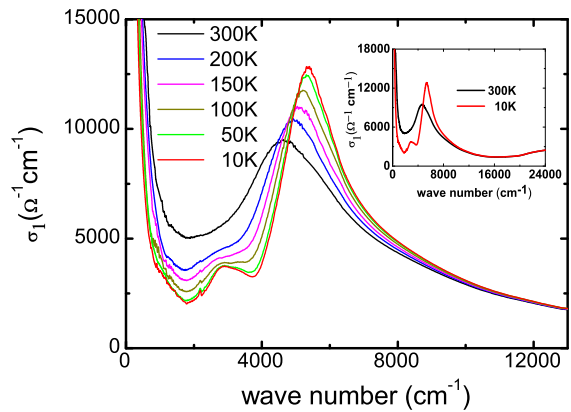}}
\caption{(Color online) The frequency dependence of the real part
of optical conductivity at different temperatures. Inset shows the
$\sigma_1(\omega)$ at 10 K and 300 K over a broad energy range.}
\end{figure}

Surprising results were observed at low temperatures. Roughly above
but near 200 K, we can identify the development of a new shoulder
near 2700 \cm in R($\omega$) and it becomes more and more dramatic
with the temperature decreasing. Furthermore, the dip near 3500 \cm
shifts to higher frequency and becomes much more pronounced than the
high temperature data. Then, in the $\sigma_1(\omega)$ spectra, a
new peak develops near 2800 \cm. Meanwhile, the spectral weight in
the mid-infrared region near this area is strongly suppressed and
transferred to the very strong peak near 5000 \cm, causing a
sizeable shift of original CDW gap towards higher energy. The
transfer of the spectral weight is clearly seen in the inset of Fig.
4. As we shall explain below, the new peak developed near 200 K is
an indication of new energy gap structure, which could be ascribed
to the formation of a new CDW order. The change of the spectral
weight at higher energies suggests that the band structure is
substantially reconstructed with the formation of the new
order\cite{ARPESCeTe3400meV}. Additionally, the free carrier
behaviors also change in response to the formation of the new CDW
gap. The reflectance edge near 2500 \cm becomes sharper than the
edge at high temperatures, reflecting a reduction of the free
carrier scattering rate.

It is also worth noting that a small peak feature near 2200 \cm in
R($\omega$) develops along with the appearance of the above
mentioned shoulder while temperature decreases, which also leads
to a peak in the $\sigma_1(\omega)$ spectra roughly at the same
frequency. Different from the notable shift of the two broad peaks
corresponding with the two CDW orders, the small peak position
changes little at varied temperatures. We noticed that it couldn't
be ascribed to the phonon mode because it locates at such high
energy which is not expected based on the atomic mass of both Ce
and Te. The origin of the peak remains to be explored.

\section{\label{sec:level2}DISCUSSION}
We shall first elaborate the change of the conducting electrons with
the development of the new energy gap. To estimate in a quantitative
way, we decompose the conductivity spectra into Drude and Lorentz
components at two representative temperatures, 300 K, the room
temperature, and 10 K, the lowest measurement temperature:

\begin{equation}
\epsilon(\omega)=\epsilon_\infty-{{\omega_p^2}\over{\omega^2+i\omega/\tau_D}}+\sum_{i=1}^N{{S_i^2}\over{\omega_i^2-\omega^2-i\omega/\tau_i}}.
\label{chik}
\end{equation}
Here, $\epsilon_\infty$ is the dielectric constant at high energy,
the middle and last terms are the Drude and Lorentz components,
respectively. The results are shown in Fig. 5. At room
temperature, the $\sigma_1(\omega)$ spectrum could be well
reproduced by two Drude components and two Lorentz terms. The two
Drude components, one with very narrow peak width and one with
much broader peak width, describe the conducting carrier responses
arising from different bands or Fermi surfaces. The Lorentz peak
structure near 4800 \cm reflects the CDW gap. At 10 K, one more
Lorentz component near 2800 \cm is added to reproduce the new CDW
gap feature. However, the broader Drude component becomes
intensively suppressed while the narrow Drude component displays a
relatively smaller change. The fit parameters for the
Drude-Lorentz model are listed in table 1. In general, we found
that both the $\omega_p$ and 1/$\tau$ decrease at lower
temperature. This result could be easily interpreted: the
formation of the new CDW gap further removes those electrons near
E$_F$ which experience stronger scattering, leading to a reduction
of conducting carrier density; meanwhile the scattering rate is
reduced due to the reduction of scattering channels. It is noted
that in the dc resistivity $\rho$(T) of CeTe$_3$, no obvious
anomaly below 300 K was observed\cite{2CDW}. This is also
understandable. From the semi-classic Boltzmann transport theory,
the resistivity is determined by the complex function of Fermi
velocity, the scattering rate, and their weighted integral over
the whole FS. Whether $\rho$(T) shows an anomaly depends on the
balance of those quantities which could experience substantial
changes across the transition.

\begin{figure}[b]
\scalebox{0.3} {\includegraphics [bb=480 20 8cm 18cm]{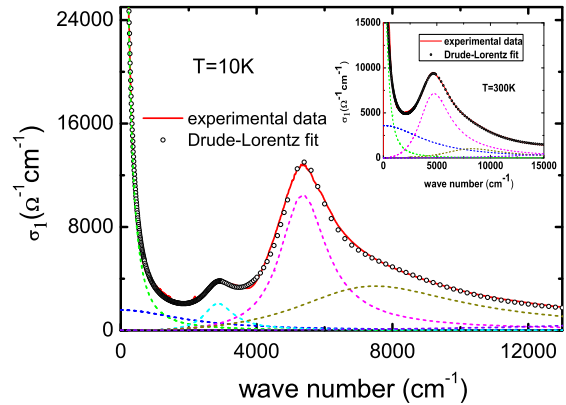}}
\caption{(Color online) The experimental data of
$\sigma_1(\omega)$ spectrum at 10 K and the Drude-Lorentz fit
result. Each term is displayed in the dashed line at the bottom.
Inset shows the result at 300 K.}
\end{figure}

The central issue of the present paper is the observation of the new
gap feature at about 2800 \cm near 200 K. As the gap formation is a
characteristic feature for a density wave instability, the structure
could naturally be ascribed to a new CDW order at low temperature.
From the analysis of the free carrier response, we found that the
new CDW order mainly occurs in the bands or Fermi surfaces
responsible for the broader Drude component. Those states were
gapped away from the Fermi level E$_F$. As we already pointed out,
not only those states were affected by the new order, the bands far
away from the E$_F$ were also influenced as well, because the gap
corresponding to the first CDW at high temperature was shifted to
higher energy, and there is a spectral weight transfer from the low
energy to high energy as well.

\begin{table*}[htbp]
\begin{center}
\newsavebox{\tablebox}
\begin{lrbox}{\tablebox}
\begin{tabular}{*{14}{m{8mm}}}

%\arrayrulewidth=1pt
\hline\\[0.4pt]
{}&$\omega_{p1}$&$\tau_{D1}$&$\omega_{p2}$&$\tau_{D2}$&$\omega_1$&$\tau_1$&$S_1$&$\omega_2$&$\tau_2$&$S_2$&$\omega_3$&$\tau_3$&$S_3$\\
\hline\\[0.4pt]
300K&27&0.42&31&4.5&--&--&--&4.7&3.5&39&8.3&6.6&20\\[4pt]
10K&27&0.17&14&2.1&2.9&1.1&11&5.4&1.6&32&7.5&5.9&35\\
\hline
\end{tabular}
\end{lrbox}
\caption{Temperature dependence of the plasma frequency $\omega_p$
and scattering rate $\tau_D$ of the Drude term, the resonance
frequency $\omega_i$, the width $\tau_i$ and the square root of
the oscillator strength $S_i$ of the Lorentz component(all entries
in 10$^3$ \cm). Two Drude terms and other two low energy Lorentz
ones are displayed at room temperature. One more Lorentz mode is
added at 10K.} \scalebox{1.0}{\usebox{\tablebox}}
\end{center}
\end{table*}

It deserves to remark that the new CDW order developed at low
temperature in CeTe$_3$ as indicated by the present measurement was
not seen by any other experimental techniques
previously\cite{ARPESCeTe3400meV,ARPESCeTe3,RTe3che.pres,CeTe3pres,magRTe3,CeTe3mulphase,2CDW}.
We emphasize here that the optical spectroscopy measurement is a
bulk probe technique. One needs to further examine carefully the
data from other measurement techniques. In fact, the two different
CDW orders were observed in a number of heavy rare-earth based
RTe$_3$ compounds with the second transitions occurring between 50
to 200 K\cite{2CDW,ARPES2CDW,opticsHoEr}, our measurement suggests
that similar second CDW transition is also present in the light
rare-earth element based compound. Even if the static CDW is not
formed, the very strong CDW fluctuations are already present.

\section{\label{sec:level2}CONCLUSIONS}
To conclude, we report an optical study of the single crystal
CeTe$_3$, a rare-earth element tri-telluride, which belongs to the
layered quasi-two-dimensional charge density wave systems. We
observed strongly temperature dependent optical spectra, which is
in sharp contrast to the early report. Besides the large CDW gap
feature which already exists at room temperature, we also observed
the development of another CDW order near 200 K. The second CDW
gap at low temperature removes the electrons near E$_F$ which
undergo large scattering. The Fermi surface is still partially
gapped with good metallic behavior exhibiting in the entire
temperature range.

\begin{center}
\small{\textbf{ACKNOWLEDGMENTS}}
\end{center}
This work is supported by the National Science Foundation of China,
the Chinese Academy of Sciences, and the 973 project of the Ministry
of Science and Technology of China.

\end{document}